\documentstyle[preprint,revtex]{aps}
\begin{document}
\draft
\begin{title}
\begin{center}
High Energy Electron Proton Scattering in View of Born Approximation
\end{center}
\end{title}
\author{M.T. Hussein and N.M. Hassan }
\begin{instit}
 Physics Department, Faculty of Science, Cairo University
\end{instit}
\begin{abstract}
 The hadronic quark structure is investigated in the frame of  high
 energy electron proton scattering. A phenomenological model based on
 the Born approximation is used to calculate the transition matrix
 element for the quark system forming the proton target. A potential
 of electromagnetic nature is assumed for the calculation of the
 multiple scattering of electron with the constituent valance quarks
 of the proton.  It is found that the first two terms of the Born
 approximation are sufficient to describe the experimental data of
 differential cross section for the electron proton system only at low
 momentum transfer square. On the other hand, a two body scattering
 amplitude with a relevant form factor may give proper agreement
 extended to high momentum transfer square region. A harmonic
 oscillator wave function is used to reproduce the data of the
 experiment at low momentum transfer. This satisfies quietly the
 confinement condition of the quarks in the hadron system. However the data
 of the experiment at relatively high momentum transfer, show that hadronic
 quarks behave more freely and may be described by just a Coulomb wave
 function. \\
\end{abstract}

\begin{tabular}{cccc}
Experiments & SLAC-E-140 & $Q^2=1-3 (GeV/c)^2$ & $E_e= 8 GeV$\\
& SLAC-E-136 & $Q^2=2.9- 31.3 (GeV/c)^2$ &$E_e= 5-21.5 GeV$\\
& SLAC-NH-11& $Q^2=1.75-8.83 (GeV/c)^2$ & $E_e= 1.5-9.8 GeV$\\
\end{tabular}
\vskip 3 cm

\narrowtext
\section{INTRODUCTION}
\label{sec:intro}
The most recent data of the electron-proton (ep) scattering experiments at
SLAC [1-3] have played a significant role in investigating the hadron
structure. In the last decade, many trails have been executed to deal with
the ep scattering problem. The Glauber multiple scattering approach [4] was
used to present an eikonal picture for the ep scattering. The quantum
chromodynamics QCD [5,6] was developed as the most convincing present
theory for the interactions between those constituent quarks, formulated in
terms of the exchange of colored vector gluons. At sufficiently high values
of $Q^2$, the running strong coupling constant is expected to become small
enough, due to the properly of asymptotic freedom, to allow the use of
perturbation theory to simplify QCD. However, there is considerable
controversy as to how large a value of $Q^2$ is sufficient for perturbative
quantum chromodynamics PQCD [7]. The exclusive processes such as electron
proton (e-p) scattering are predicted to have a single dimensional scaling
[8] at large $Q^2$. In this case only the valance quark states are important
and a rough idea of the $Q^2$ dependence can be gained by simply counting the
number of quark-gluon vertices. Elastic form factor, for example, should
scale asymptotically as $(Q^2)^{-(n-1)}$, where n is the number of valance quarks
participating in the interaction. For (e-p) scattering n=3 thus the
structure function behaves as $~ q^{-4}$. Perturbative QCD [7] predicts
calculable logarithmic departures from the $Q^2$ dependence of exclusive
amplitudes given by the simple dimensional scaling law. The earliest
efforts [9] used unrealistic symmetric distribution amplitudes and required
a large multiplicative factor to normalize the results to the data at $Q^2$
$\approx 10 (GeV/c)^2$.  Chernyak and Zhitnitsky [10,11] proposed a model form for
the nucleon distribution amplitude which satisfies the sum rules and in
which the momentum balance of the valance quarks in the proton is quite
asymmetric. The result of this model is justified in the diquark model
[12,13]. The nonperturbative calculations succeeded in modeling the
region $Q^2 < 20 (GeV/c)^2$ fairly well but some difficulties have been found at
large values of $Q^2$. By this article we aim to develop the eikonal optical
picture of the (ep) scattering on the bases of the Born multiple scattering
of the incident electron with the constituent valance quarks of the target
hadron. The paper is organized so that in section 2, we present the
hypothesis and the mathematical formalism of the model. Results and
discussion as well the comparison with experimental data are given in
section 3.

\section{The Model}
  During the electron-proton (e-p) scattering, we consider the proton as a
  bag  including three valance quarks. We proceed using the semi-classical
  Born approximation which is a useful technique when the de Broglie
  wavelength of the incident particle is sufficiently short compared with
  the distance in which the potential varies appreciably. In this
  approximation we consider the multiple scattering series in which the
  projectile interacts repeatedly with the potential and propagates freely
  between two such interactions. We expect that the Born series converges
  if the incident particle is sufficiently fast so that it cannot interact
  many times with the potential and/or if the potential is weak enough.
  Starting with a plane wave as a zero order wave function of the incident
  electron, and by successive iteration, the nth order of the Born
  scattering amplitude is given by,
\begin{equation}
 \bar f_n=-2 \pi^2 <\phi_{k_f} |U|\Psi_{n-1}>
\end{equation} 
where $f_k$, is the zero order plane wave and,
\begin{equation}
\Psi_n=\phi_{k_f}(\bar r)+\int G_0(\bar r,\bar r') U(\bar r) \Psi_{n-1}(\bar r') d\bar r'
\end{equation} 
and 
\begin{equation}
G_0(\bar r,\bar r')= {exp(i k |r-r'|) \over |r-r'|}
\end{equation} 
is the free Green's function. 
Let us denote by $\bar r_o, \bar r_1,\bar r_2$ and $\bar r_3$ the coordinates 
of the incident electron and the constituent valance quarks of the target 
proton. According to the first Born approximation, the scattering amplitude is 
given by:
\begin{equation}
  f_1=-(2\pi)^2 \int e^{i \bar q.\bar r_o} A(r_o) d\bar r_o
\end{equation} 
     	
\begin{equation}
  f_1=-(2\pi)^2 {4\pi \over q} \int sin(q r_o) A(r_o) d\bar r_o
\end{equation} 
This term represents the two-body scattering amplitude or the impulse
approximation. The second Born approximation is,
\begin{equation}
  f_2=-(2\pi)^2 \int <k_f|A|\kappa> {1\over \kappa^2 -k^2 -i \epsilon} <\kappa|A|k_i> d\kappa
\end{equation} 
which represents the double scattering term in the scattering amplitude.
A($r_o$) is the average potential acting on the incident electron due to the
target valance quarks {ri} with initial and final states $\Psi_a$ and $\Psi_b$
so that,
\begin{equation}
A(r_o)=\int  {\Psi_b}^*(\{r_i\}) V(\{r_i\}) {\Psi_a}(\{r_i\}) \{d\bar r_i\}
\end{equation} 
In the present situation we consider that the electron-quark (eq) potential
to be a pure Coulomb with the form,
\begin{equation} 
V(\{r_i\})={1\over r_{o1}}- {1\over r_{o2}}- {1\over r_{o3}}
\end{equation} 
 	
and $r_{oi}$ is the relative coordinate  $r_{oi} = r_o - r_i$
The term  $1/r_{oi}$ may be expanded in terms of the spherical harmonics $Y_{l,m}$ as

\begin{equation} 
{1\over r_{oi}}=\sum_{l=0} {\sum_{m=-l}}^l {4\pi\over 2l+1}{(r_<)^l \over (r_>)^{l+1}} Y_{l,m}(\hat r_o) Y^*_{l,m}(\hat r_i)
\end{equation} 

where $r_> (r_< )$ is the greater (the lesser) of $r_o$ and $r_i$. The radial
integral of Eq.(3) may be divided into two parts:

\begin{equation} 
\int ^{\infty}_0 =\int ^{r_o}_0 +\int ^{\infty}_{r_o} 
\end{equation} 
 										(10)
and substitute for $r_> (r_< )$ by $r_i (r_o)$ in the first integral and 
for $r_> (r_<)$ by $r_o (r_i) $
in the second integral. Then writing
\begin{equation} 
A(r_o)=\sum _{i=1}^3 A(r_i)
\end{equation} 

where $A(r_i)$ is the average potential due to the $i^th$ quark, then
\begin{equation} 
A_i (r_o)=\int{\Psi_b}^*(\{r_i\}) {1\over r_{oi}} {\Psi_a}(\{r_i\}) \{d\bar r_i\}
\end{equation} 

$$A_i (r_o)=\int{\Psi_b}^*(r_i) {1\over r_{oi}} {\Psi_a}(r_i) d\bar r_i $$
\begin{equation} 
= 4\pi \int^{r_o}_0 |\Psi (r_i)|^2 {1\over r_o} r_{i}^2 dr_i 
+ 4\pi \int^{\infty}_{r_o} |\Psi(r_i)|^2  r_i dr_i 
\end{equation} 

The evaluation of $A_i(r_o), \{i=1,2 and 3\}$ for the three quark system 
differs only in their signs due to the charge situation. On the other and the
differential cross section is calculated either in terms of the multiple
scattering Born approximation amplitude,
\begin{equation} 
d\sigma /d\Omega= |f_1+f_2+ ...|^2 
\end{equation} 

or in terms of the two body scattering amplitude with a proton form factor
correction.

\begin{equation} 
d\sigma /d\Omega= |f_o|^2  F^2 (q)
\end{equation} 
here, $f_o$ is the scattering amplitude of the electron scattering by a
point-like proton. The proton form factor F(q) is also an important
physical quantity which reflects information about the particle structure.
It is defined as:
$$F(q)= < O'|e^{i \bar q.\bar r}|O > $$
\begin{equation} 
F(q)=2\pi \int \Psi' (\bar r) e^{i q r cos(\theta)} \Psi (\bar r) r^2 d cos(\theta) dr
\end{equation} 

as the model is a phenomenological type, it is then branches in two ways,\\
I) The first one assumes that quarks are confined inside the bag by an
extremely deep potential, and the quark states are reasonably represented
by a harmonic oscillator wave function.\\
II) The second way assumes that the constituent quarks behave as almost
free particles as is suggested by experiments of deep inelastic scattering
[1,14]. However, in all cases the $(e^-p)$ scattering is treated as the
collision between the high energy incident electron with a composite proton
system. Proceeding with the first assumption ( I ) and assuming a harmonic
oscillator wave function for the valance quarks,

\begin{equation} 
\Psi_a = \Psi_b =({2\over \pi a^2})^{3/4} e^{-r^2 / a^2} 
\end{equation}

Then the average scattering potential is;
\begin{equation} 
A(r_o)=4\pi [{(2\pi)^{3/2} \over a e^{2r^2/a^2}}+ 
{\Gamma (3/2,0,r^2/a^2)\over 2\pi^{3/2} r}]   
\end{equation} 
where $\Gamma (3/2,0,r^2/a^2)$ is the incomplete Gamma function of order 3/2. 
The scattering amplitude and the differential cross section are to be
calculated numerically. Moreover, the form factor is calculated for the
proton system as,

\begin{equation} 
F_{h_osc}(q)=1 /exp(a^2 q^2 /8)^2
\end{equation} 
Considering now the second case (II) of elastic scattering and that all
quarks are in the 1S ground state of a Coulomb wave function,

\begin{equation} 
\Psi_a = \Psi_b ={1\over \sqrt{\pi a^3}} e^{-r / a} 
\end{equation} 
'a' is the proton radius , then
\begin{equation} 
A(r_o)={1\over r_o}+{a+2r_o \over a^2 e^{-2r_o /a}} -{ a+2 r_o+2r_{o}^2\over a^2 e^{-2r_{o}/a}} 
\end{equation} 
Consequently,
\begin{equation} 
f_1(q)=[{\pi\over 2q} - {a\over4+a^2 q^2} -{1\over q} atn (a~ q/2)]
\end{equation} 
\begin{equation} 
d\sigma /d\Omega =4 [{\pi\over 2q} - {a\over4+a^2 q^2} -{1\over q} atn (a~ q/2)]^2
\end{equation} 
And the form factor is then,
\begin{equation} 
F_{Coul}(q)= 16 /(4+a^2 q^2)^2
\end{equation} 

\section{Results and discussion}
Data used in this article are those from the experiments coded SLAC-E-140  
[1] at low momentum transfer extended from 1 to 3 $(GeV/c)^2$ using electron 
beam of energy 8 GeV, and SLAC-E-136 [2] at a wide range extended from 2.9 to 
31.3 $(GeV/c)^2$. The later was conducted using accelerated electrons with 
energies from 5 to 21.5 GeV were elasticaly scattered by protons in a liquid
-hydrogen target at Stanford Linear Accelerator Center. The center of mass 
energy of the reaction is found in terms of the energies $( E_e \& E_p)$ of 
the scattered electron and proton respectively and their opening angle $\theta$. 

\begin{equation} 
\sqrt s=\sqrt {m_{e}^2+ m_{p}^2+4 E_e E_p sin^2(\theta /2)}
\end{equation} 

The differential cross section is calculated in the frame of multiple scattering 
using Eq.(14) for both the experiments SLAC-E-136 and SLAC-E-140. The result of 
calculation as well as the experimental data are displayed in Figs.(1) and (2) 
respectively. It is found that only the first two terms of the Born  series give 
appreciable contribution to the differential cross section. The second term 
contributes for not more than 10 $\%$ in all cases. The calculations are carried 
out using harmonic oscillator wave function (solid lines) and Coulomb wave function 
(dashed lines) for the valance quarks forming the proton system. The experimental 
data are reproduced to a limited extend by the harmonic oscillator wave function 
only at low momentum transfer region. On the other hand the use of the Coulomb 
wave function shows better representation of the data at high momentum transfer. 
This result reflects the fact that quarks inside the hadron are quite confined 
at low momentum transfer region while at high momentum transfer, quarks behave 
more freely and the confinement condition becomes no longer necessary. However, 
the comparison with the experiment shows also that the multiple scattering 
approach couldn't give satisfactory agreement particularly at high momentum 
transfer regions. The differential cross section is recalculated in terms of 
the proton form factor according to Eqs.(15-16). Proceeding the same analogy, 
so that the proton form factor is calculated twice 
using the harmonic oscillator and the Coulomb wave functions. The result are 
displayed in Fig.(3) compared with data of the experiment SLAC-NH-11 [3] as 
well as the prediction of the simple dipole model of the following form,
\begin{equation}
F_{SD}= {1 \over  (1+ q^2 / 0.71)^2}
\end{equation}
The prediction of Eq. (15) is displayed with the experimental data in Figs.
(4) and (5) for the experiments E-136 and E-140 respectively. It is clear 
now that the two body scattering amplitude corrected with a relevant form 
factor may reproduce the experimental data quite well, assuming harmonic 
oscillator wave function for the constituent quarks. In the above 
calculations, we use the units where $(\hbar =c=1)$, hence the momentum in 
GeV/c may be represented in units of $ fm^{-1} $ with a conversion factor 0.2 
In all the above cases the potential parameters are determined by the Chi-
Square fitting method. A root mean square radius of the proton is found to be 
0.8 and 1.0 fm on using the harmonic oscillator and the Coulomb wave functions 
respectively. The distribution of the hadronic system produced in e-p scattering 
is measured by ZEUS collaboration [15] at center of mass energy of 296 GeV. 
Comparison of the results with the QCD radiation has a strong influence on the 
characteristics of the final state. The data are reasonably reproduced by the 
Lund model based on a matrix element calculated in the first order of  $\alpha$, 
followed by a appropriate parton shower, as well as by the color dipole model. 
The HERWING parton shower model also gives a reasonable representation of the 
data. Neither the first order matrix elements alone nor the Lund parton shower 
model, without the matrix element calculation.

\figure{The differential cross section of ep scattering as calculated by the multiple scattering approach
using for the proton valance quarks a harmonic oscillator wave function (solid line), and the 
Coulomb wave function (dashed line). The experimental data of SLAC-E-136 are represented 
by the cross signs (x).}
\figure{The differential cross section of ep scattering as calculated by the multiple scattering approach
using for the proton valance quarks a harmonic oscillator wave function (solid line), and the 
Coulomb wave function (dashed line). The experimental data of SLAC-E-140 are represented 
by the cross signs (x).}
\figure{The proton form factor calculated using for the proton valance quarks a harmonic oscillator
wave function (solid line), and the Coulomb wave function (dashed line). The prediction of the  
simple dipole model is represented by (dotted line). The experimental data of SLAC-NE-11 are 
represented by the cross signs (x).}
\figure{The differential cross section of ep scattering  calculated as two body scattering amplitude
corrected with a relevant form factor using for the proton valance quarks a harmonic oscillator 
wave function (solid line), and the Coulomb wave function (dashed line). The experimental 
data of SLAC-E-136 are represented by the cross signs (x).}
\figure{The differential cross section of ep scattering  calculated as two body scattering amplitude
corrected with a relevant form factor using for the proton valance quarks a harmonic oscillator 
wave function (solid line), and the Coulomb wave function (dashed line). The experimental 
data of SLAC-E-140 are represented by the cross signs (x).}

\begin{references}
\bibitem{}  A.F.Sill et.al., Phys.Rev.D48, 29 (1993).
\bibitem{} P. Bosted et al., Phys. Rev. Lett. 68, 3841 (1992).
\bibitem{} R.C.Walker, Phys. Lett. 224B, 353 (1989).
\bibitem{} M.T. Hussein, 24th International CosmicRay Conference ICRC, Roma, HE,135  (1995).
\bibitem{} N.G.Stefanis, Phys.Rev. D40, 2305 (1989).
\bibitem{} C.E.Carlson, M. Gari and N.G.Stefanis, Phys.Rev.Lett. 58, 1308 (1987).
\bibitem{} -S.J.Brodsky and G.P.Lepage, Phys.Rev.D22, 2157 (1980).\\
    -G.P.Lepage and S.J.Brodsky, Phys.Rev.Lett. 43,545 (1979). \\
    -A.V.Radyushkin, Nucl.Phys. A532, 141c (1991).
\bibitem{} S.J.Brodsky and G.Farrar, Phys.Rev.D11, 1309 (1975).
\bibitem{} A.Duncan and A.H.Muller, Phys.Lett 90B, 159 (1980).
\bibitem{} V.L. Chernyak and I.R. Zhitnitsky , Nucl.Phys. B246, 52 (1984).
\bibitem{} M.Gari and N.G. Stefanis, Phys.Lett. B175, 462 (1986).
\bibitem{} Z.Dziembowski and J.Frankil, Phys.Rev.D42, 905 (1990).
\bibitem{} P.Kroll, M. Schurmann and W. Schweiger, Z.Phys. A338, 339 (1991).
\bibitem{} R.G.Arnold et.al., Phys.Rev.Lett.57, 174 (1986).
\bibitem{} ZEUS Collaboration, Z. Phys. C59, 231 (1993).

\end{references}
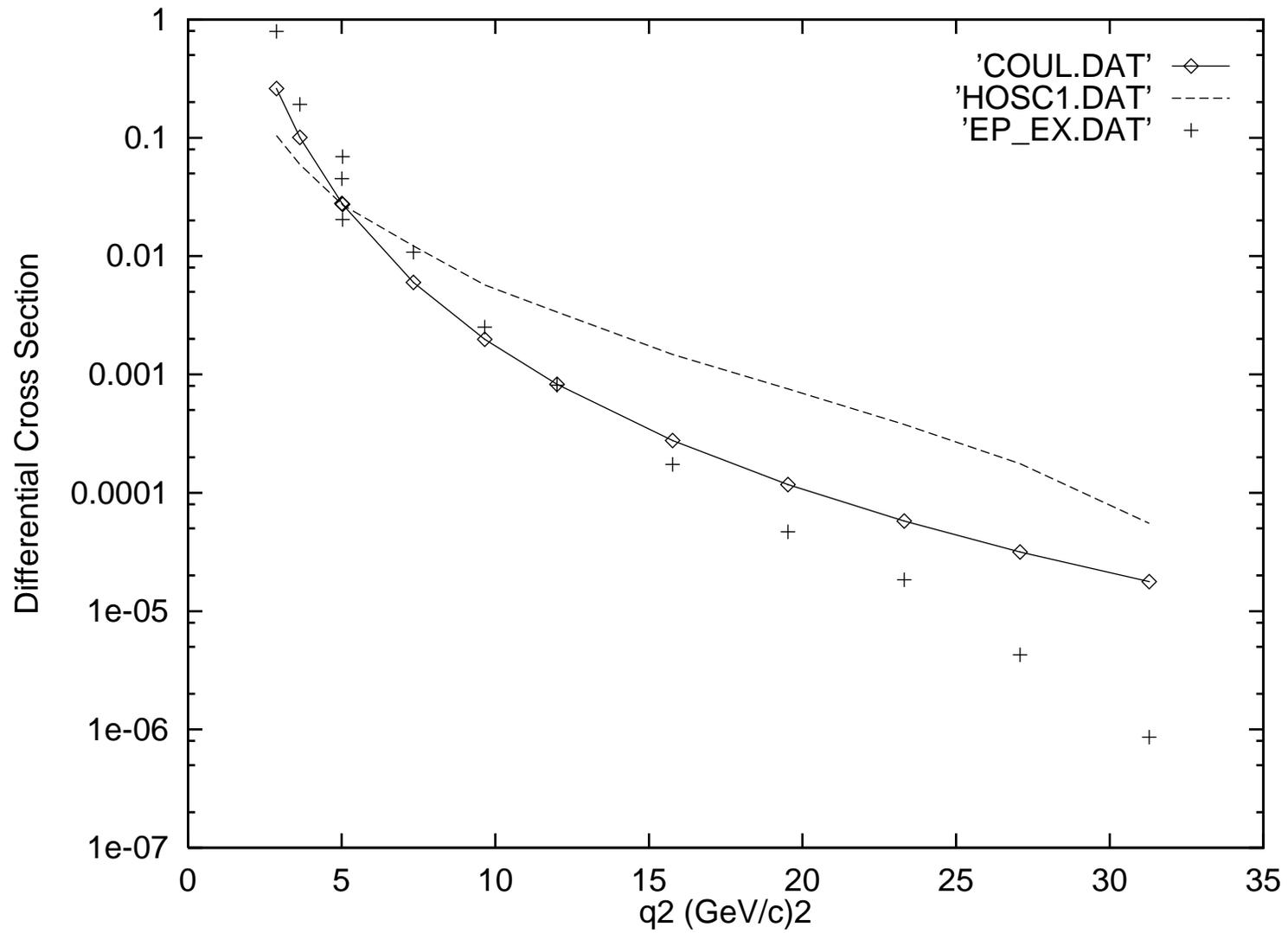
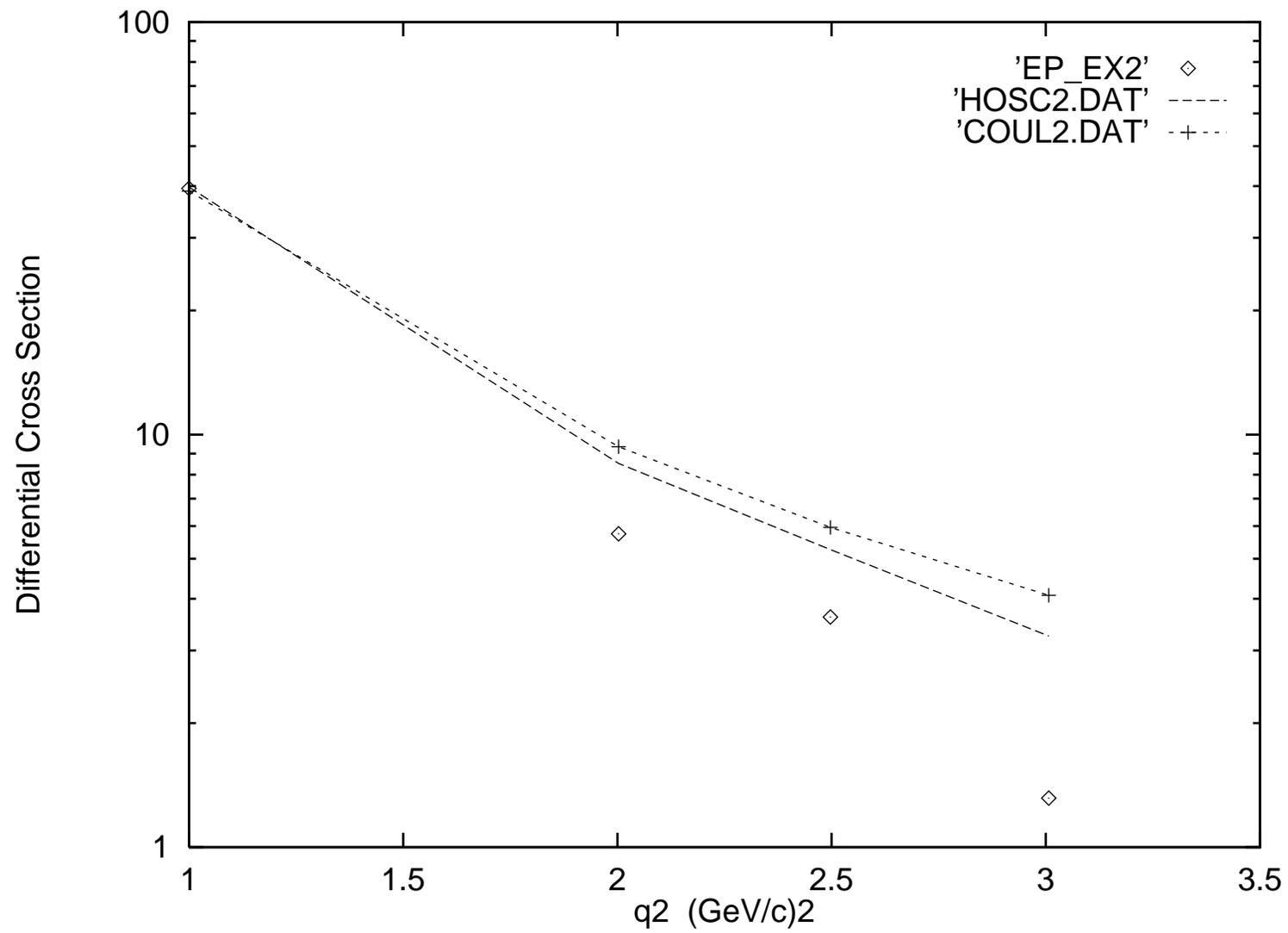
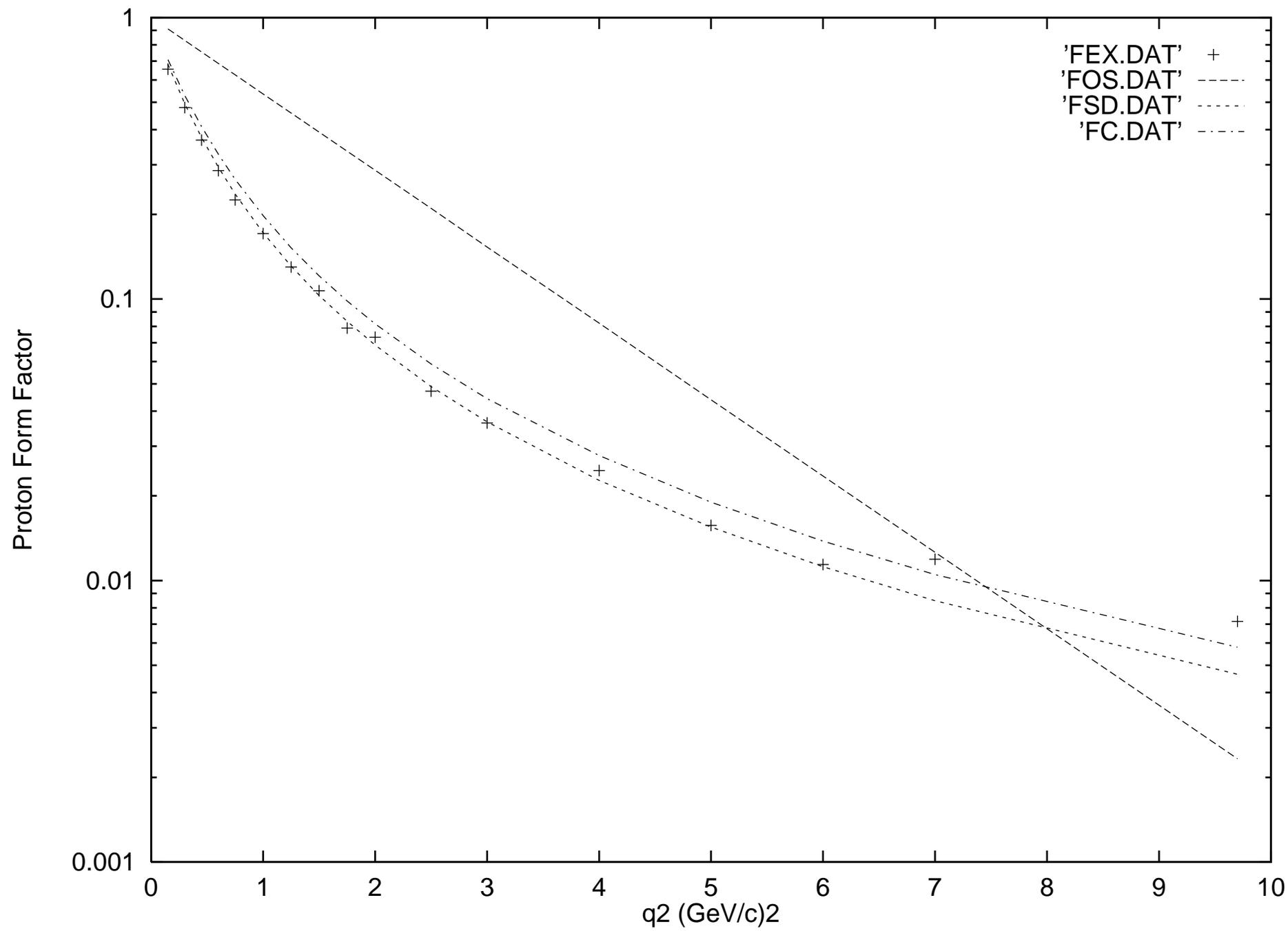
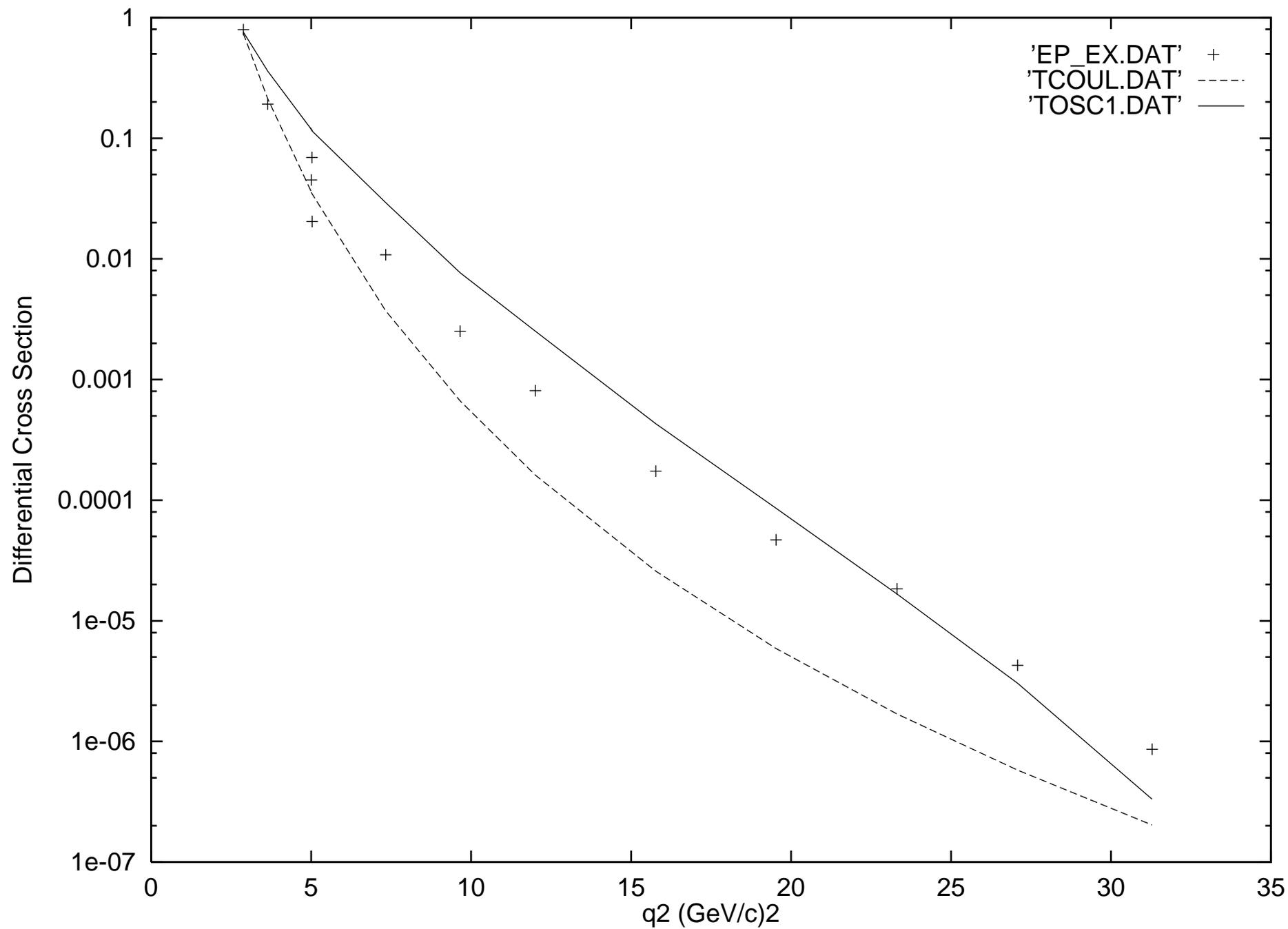
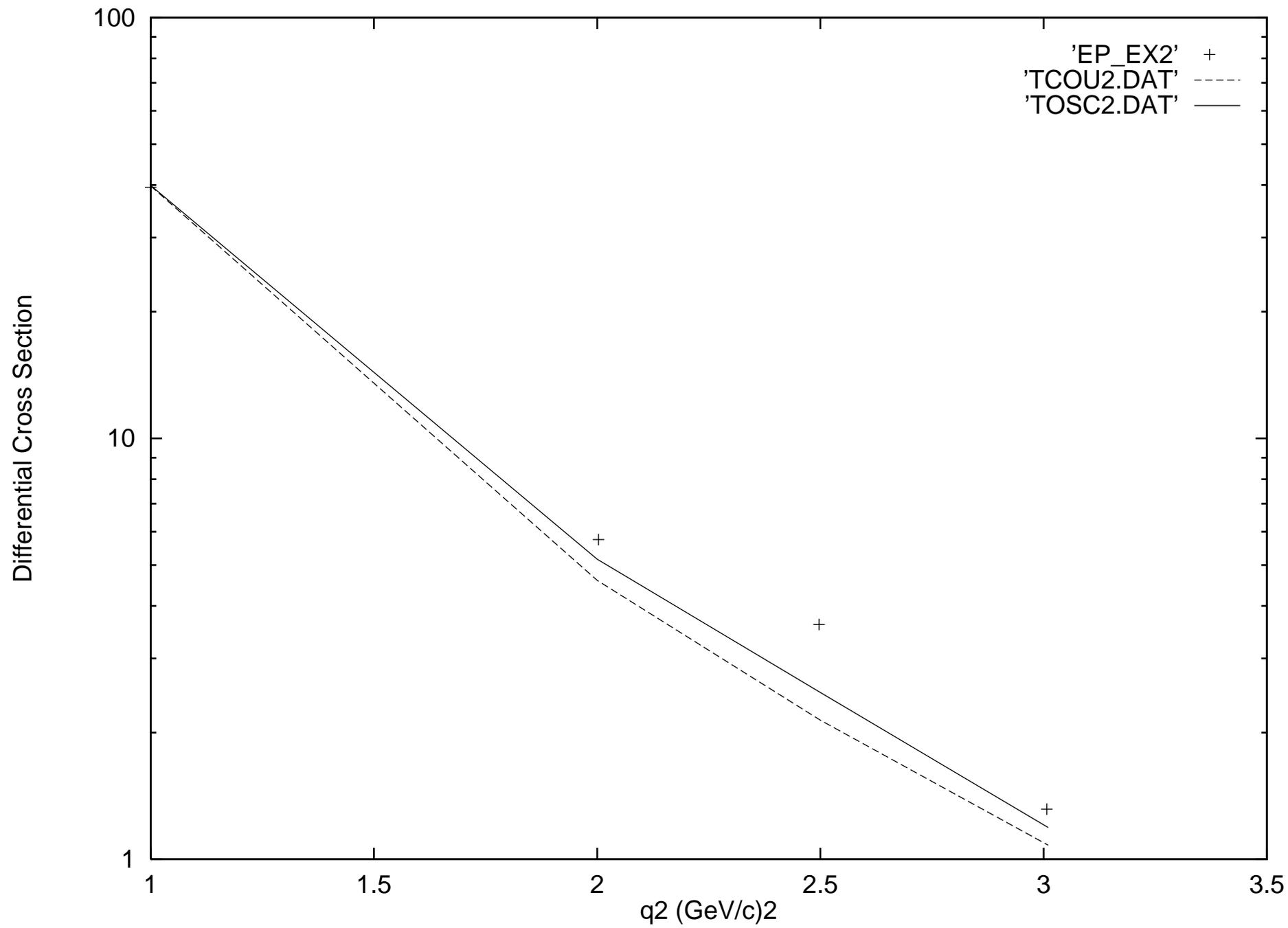
\end{document}